# RIEMANN SURFACE AND QUANTIZATION

Perepelkin E.E., Sadovnikov B.I., Inozemtseva N.G.

**Abstract**

This paper proposes an approach of the unified consideration of classical and quantum mechanics from the standpoint of the complex analysis effects. It turns out that quantization can be interpreted in terms of the Riemann surface corresponding to the multivalent $\operatorname{Ln}\Psi$ function. A visual interpretation of «trajectories» of the quantum system and of the Feynman's path integral is presented. A magnetic dipole having a magnetic charge that satisfies the Dirac quantization rule was obtained.

**Introduction**

Considering the microcosm effects quantum mechanics occupies a special place having many formulations [1]. The issues associated with the scope of applicability and the transition from classical mechanics to the quantum one and vice versa are significant [2-5]. Terms such as «orbit», «quantization of orbits and energy», path integral (Feynman) [6,7], and many others are introduced quite arbitrarily, appealing mainly to an experiment.

In the present work these terms are considered from the point of the complex analysis [8,9]. Such consideration gives a visual interpretation of pure quantum effects in terms of classical (deterministic) concepts.

For example, the absence of a specific trajectory of a quantum particle from the position of the complex analysis means that a quantum particle is not a point (as it is in classical mechanics), and it occupies some domain $D$ in a complex plane. The result of the quantum particles movement (quantum object) is that domain $D$ becomes $D'$. So, there is a mapping of one domain to another. In this case each point $\xi \in D$ goes into some point $\xi' \in D'$ along some complex path $\xi(t)$. Thus, the quantum system at the same time «moving» along an infinite set of trajectories $\xi(t)$ that leads to a well-known formulation of the Feynman's path integral.

Another example is the quantization of the energy levels and the radii of the orbits in the atom. From the point of the complex analysis emergence of quantization is related to the concept of Riemann surface [10-13], corresponding to the multivalent $\operatorname{Ln}\Psi$ function. Each orbit in an atom can be supplied with a corresponding Riemann surface. At the origin (in the center of atom) there is a pole of the flow rate probabilities function of the first order $\langle \vec{v} \rangle = \frac{\hbar k}{r} \vec{e}_\phi$. As a result, bypassing the circuit at the origin of coordinates (Bohr-Sommerfeld quantization rule), there is a movement to the next leaf of the Riemann surface $\frac{\tilde{S}}{\hbar} = \varphi \sim 2\pi k, k \in \mathbb{Z}$. The increase valent number $k$ leads to a transition into classical mechanics $\frac{\tilde{S}}{\hbar} = \varphi \gg 1$.

As a fundamental concept of quantum mechanics is the wave function, which is a complex function, then the mathematical apparatus of complex analysis is natural for quantum mechanics. Note that the basic equations of quantum mechanics, such as the Schrödinger equation and the Pauli one, are directly connected with the continuity equation [14] for the density function of probability $f(\vec{r},t) = |\Psi|^2$ and velocity of the probability flow $\langle \vec{v} \rangle (\vec{r},t)$. This connection allows one to use a complex approach not only in quantum mechanics but also in classical mechanics and hydro-dynamics. It also works in reverse, that is, the use of classical hydro-dynamic solutions in quantum mechanics. From here we can see the possibility of the existence of quantum shock wave systems, to which many studies are devoted to [34-41].

The paper has the following structure. §1 provides the basic formulas derived in [14]. It is shown that the phase $\varphi$ of the wave function $\Psi$ can be interpreted in two ways. On the one hand as action $\tilde{S}$, and on the other hand as the scalar velocity potential $i\Phi = \mathrm{Ln}\left(\frac{\Psi}{\bar{\Psi}}\right)$ of the probabilistic flow $\langle \vec{v} \rangle (\vec{r},t)$. The connection between classical and quantum potential energy [15-17], which is included into the Schrödinger and Pauli equation [2-5].

In §2 the wave function is regarded as a conformal mapping $\Psi = e^Z$ of some complex domain $D_Z$ to domain $D_\Psi$. The value $Z = \frac{1}{2}(S + i\Phi)$, where $S = \mathrm{Ln}(\Psi\bar{\Psi})$. Mapping $\Psi = e^Z$ is univalent in $0 < \Phi < 4\pi$. Thus it is shown that Jacobian $J$ of mapping $\Psi = e^Z$ is a probability density $J = \frac{1}{4}|\Psi|^2$. That is, it satisfies the continuity equation with velocity vector field $\langle \vec{v} \rangle (\vec{r},t)$. The corresponding inverse mapping $Z = \mathrm{Ln}\,\Psi$ is multivalent complex function that leads to the interpretation of ratio $\frac{\tilde{S}}{\hbar} = \frac{\Phi}{2} \sim 2\pi k$ as the number $k$ of leaves of the Riemann surface of function $\mathrm{Ln}\,\Psi$. Increasing the valent number $k$ ( $\frac{\tilde{S}}{\hbar} \gg 1$ ) there is a «transition» into classical mechanics. With small valent numbers $k$, for example $\frac{\tilde{S}}{\hbar} \sim 2\pi$, (an univalent mapping) there is a «transition» into quantum mechanics.

§3 considers a complex principle of least action. The quantum system in the complex plane is represented by domains $D_Z$ and $D_\Psi$ which are connected with direct and reverse mappings $\Psi = e^Z$ and $Z = \mathrm{Ln}\,\Psi$, respectively. As a result of the quantum system evolution there is a change in domains $D_Z$ and $D_\Psi$. As a result domain $D_{\Psi_1}$ corresponding to $t_1$ transforms into $D_{\Psi_2}$ corresponding to $t_2$. Each point of $D_\Psi$ passes *on its «trajectory»* $\xi(t)$

determined by a minimum action $Z_{12} = -\frac{1}{2}\int_{t_1}^{t_2} Q d\tau + i\frac{1}{\hbar}\int_{t_1}^{t_2} \tilde{L} d\tau$, where $Q = \text{div}_r \langle \vec{v} \rangle$, and $\tilde{L}$ is the Lagrangian function. Depending on the type of gauge (Lorenz, Coulomb), we can get different conditions on the dependence of potential energy on time. In the result, the quantum system simultaneously moves along an infinite set of trajectories that leads to the formulation of the principle of least action via the path integral $\tilde{\Psi}_{12} = \int_1^2 |\Psi_{12}(\xi)| e^{i\frac{\tilde{S}_{12}[\xi]}{\hbar}} D\xi$, where $\int_1^2 D\xi$ – a conditional note of the infinitely divisible functional integration over all trajectories $\xi(\tau)$, transforming domain $D_{\Psi_1}$ into domain $D_{\Psi_2}$. In case when the probability density is conserved along the trajectories (characteristics), we can get the Feynman's path integral $\tilde{\Psi}_{12} = \int_1^2 e^{i\frac{\tilde{S}_{12}[\xi]}{\hbar}} D\xi$ [6,7].

In §4 evolution operator, which produces a mapping of domain $D_{\Psi_1}$ to domain $D_{\Psi_2}$, and having a form of $\Psi_{12}[\xi] = \text{T} e^{-\frac{i}{\hbar}\int_{t_1}^{t_2} \tilde{H} d\tau}$ is constructed. In the particular example discussed in §5, the evolution operator is a rotation in the complex plane by the angle $\Delta\varphi = \varphi_2 - \varphi_1$ transforming $D_{\Psi_1}$ into $D_{\Psi_2}$.

§5 shows that from the point of view of the complex analysis the Bohr-Sommerfeld principle of quantization [18-20] is a consequence of multivalent function $Z = \text{Ln}\,\Psi$. Since the velocity $\langle \vec{v} \rangle$ is represented as the gradient of a scalar potential $i\Phi = \text{Ln}\left(\frac{\Psi}{\overline{\Psi}}\right)$, then the value of the integral along the contour that contains the pole, $\oint_C (\langle \vec{p} \rangle, d\vec{r})$ is connected with the residue of the expanded function in the pole. While integrating along the $k$-leaves Riemann surface, the pole is passed $k$ - times, which leads to value $2\pi\hbar k$.

§6 considers examples of the vortex velocity field of the flow probabilities $\langle \vec{v} \rangle = \frac{\hbar k}{r}\vec{e}_\phi$. In section 6.1 we considered the case of the non-smooth potential $\Phi(x,y) = 2\varphi(x,y) = 2k\phi$, where $k \in \mathbb{Z}$, $\phi$ is the polar angle and $\varphi$ is the phase of the wave function, the flow of probability velocity potential and action. In this case, the momentum is $\langle \vec{p} \rangle = -i\hat{\text{p}}\varphi = \frac{\hbar k}{r}\vec{e}_\phi$. The continuity equation for the density function of the probability takes the form $(\langle \vec{v} \rangle, \nabla_r f) = 0$, that is $f = f(r)$, and the solution is obtained by the method of characteristics. Characteristics are concentric circles with the center at the pole of the function $\langle \vec{p} \rangle$ (origin of coordinates). The angular momentum is directed perpendicular to the plane of rotation $XOY$ it is quantized

$M = \hbar k$, $k \in \mathbb{Z}$. The evolution operator in this case is $e^{i\Delta\phi}$, that is, it rotates in the complex plane domain $D_\Psi$, which puts the domain $D_\Psi$ into itself.

In paragraph 6.2 we considered a model of Dirac strings. Since the vortex field (in cylindrical coordinates) $\langle \vec{v} \rangle = \dfrac{\hbar k}{r} \vec{e}_\phi$ according to [14] can be represented in the form $\gamma \vec{A}$, where $\vec{A}$ is the vector potential, then we can calculate the magnetic induction vector $\vec{B} = \operatorname{rot} \vec{A}$. It turns out that in this case, the magnetic field can be represented as $\vec{B} = -\dfrac{q_m^{(\text{Wb})} \delta(r)}{2\pi r} \vec{e}_z$ or $\vec{H} = -\dfrac{q_m^{(\text{A·m})} \delta(r)}{2\pi r} \vec{e}_z$ where $q_m^{(\text{Wb})} \stackrel{\text{det}}{=} \dfrac{2\pi \hbar k}{q_e}$, $q_m^{(\text{A·m})} \stackrel{\text{det}}{=} \dfrac{2\pi \hbar k}{q_e \mu_0}$. Value $q_m$ corresponds to a magnetic charge of Dirac [33], for which there is the quantization condition $\dfrac{q_e q_m^{(\text{Wb})}}{2\pi \hbar} = \dfrac{q_e q_m^{(\text{A·m})}}{2\pi \varepsilon_0 \hbar c^2} = k$ where $k \in \mathbb{Z}$. The obtained magnetic field has only one field line coinciding with axis OZ. Note that this system is a dipole that is not a Dirac monopole.

## § 1 The original data

Let us consider the first equation in the Vlasov chain of equations (the continuity equation) for the density function of probability distribution $f(\vec{r}, t)$ [21-25]:

$$\frac{\partial f(\vec{r}, t)}{\partial t} + \operatorname{div}_r \left[ f(\vec{r}, t) \langle \vec{v} \rangle (\vec{r}, t) \right] = 0, \tag{1}$$

where $\langle \vec{v} \rangle (\vec{r}, t)$ is the velocity of the probabilistic flow. Equation (1) can be rewritten in the form of

$$\frac{dS}{dt} = -Q, \tag{2}$$

where

$$Q \stackrel{\text{det}}{=} \operatorname{div}_r \langle \vec{v} \rangle, \quad S \stackrel{\text{det}}{=} \operatorname{Ln} f.$$

Note that when $Q = 0$, equation (1) and (2) transform into the following equation

$$f_t + \langle \vec{v} \rangle \nabla_r f = 0. \tag{2A}$$

In [14] we have established a strict relationship between equation (1) and Schrödinger and Pauli one. At the same time $f(\vec{r}, t) = \Psi(\vec{r}, t) \overline{\Psi}(\vec{r}, t) = |\Psi(\vec{r}, t)|^2$, and $\langle \vec{v} \rangle (\vec{r}, t)$ according to the Helmholtz theorem the following view was obtained:

$$\langle \vec{v} \rangle (r, t) = -\alpha \nabla_r \Phi(r, t) + \gamma \vec{A}(r, t) = \langle \vec{v}_p \rangle (r, t) + \langle \vec{v}_s \rangle (r, t), \tag{3}$$

where

$$\langle \vec{v}_p \rangle = -\alpha \nabla_r \Phi,$$
$$\langle \vec{v}_s \rangle = \gamma \vec{A}.$$
(4)

and

$$\langle \vec{v} \rangle (\vec{r},t) = i\alpha \nabla \operatorname{Ln}\left[\frac{\Psi}{\overline{\Psi}}\right] + \gamma \vec{A},$$
(5)

where $\nabla \operatorname{Ln}\left[\frac{\Psi}{\overline{\Psi}}\right] = i\nabla \Phi(\vec{r},t)$ – corresponds to the laminar flow, $\vec{A}$ – a vortex component of the velocity of the probability flow or the vector potential, that is the magnetic induction $\vec{B} = \operatorname{rot} \vec{A}$ [14]. Constants $\alpha$ and $\gamma$ in the case of consideration of a quantum system are respectively equal to $-\frac{\hbar}{2m}$ and $-\frac{q_e}{m}$ (in the particular case $-\frac{e}{m}$).

Thus, equation (1) is transformed into the equation:

$$\frac{i}{\beta}\frac{\partial \Psi}{\partial t} = -\alpha\beta\left(\hat{p} - \frac{\gamma}{2\alpha\beta}\vec{A}\right)^2 \Psi + \frac{1}{2\alpha\beta}\frac{|\gamma\vec{A}|^2}{2}\Psi + U\Psi$$
(6)

where $\hat{p} = -\frac{i}{\beta}\nabla$, $\beta = \frac{1}{\hbar}$, and the potential $U$ has the form:

$$U(\vec{r},t) = -\frac{1}{\beta}\left\{\frac{\partial \varphi(\vec{r},t)}{\partial t} + \alpha\left[\frac{\Delta\sqrt{f(\vec{r},t)}}{\sqrt{f(\vec{r},t)}} - |\nabla\varphi(\vec{r},t)|^2\right] + \gamma(\vec{A},\nabla\varphi)\right\},$$
(7)

where $\varphi(\vec{r},t)$ is the phase of the wave function $\Psi(\vec{r},t)$, which is directly connected with the scalar potential $\Phi(\vec{r},t)$ of the velocity of the probability flow (3) as

$$\operatorname{Arg}\left[\frac{\Psi(\vec{r},t)}{\overline{\Psi}(\vec{r},t)}\right] = 2\varphi(\vec{r},t) + 2\pi k = \Phi(\vec{r},t),$$
(8)

as

$$\operatorname{Ln}\left[\frac{\Psi}{\overline{\Psi}}\right] = \ln\left|\frac{\Psi}{\overline{\Psi}}\right| + i\operatorname{Arg}\left[\frac{\Psi}{\overline{\Psi}}\right] = i\Phi(\vec{r},t).$$
(9)

In the case where the probability flow has vortex component equation (6) becomes the well-known Pauli equation for particles with spin. If a vortical flow component $\vec{A}$ of the probabilities flow is absent, equation (6) becomes the classical Schrödinger equation for a scalar particle.

In [14] the following equation was obtained:

$$\frac{\partial \Phi}{\partial t} = -\frac{2}{\hbar}\left[\frac{m|\langle \vec{v} \rangle|^2}{2} + e\chi\right] = -\frac{2}{\hbar}W(\vec{r},t),$$
(10)

where $e\chi$ is the potential energy, $T = \dfrac{m|\langle\vec{v}\rangle|^2}{2}$ – kinetic, and $W(\vec{r},t)$ – total energy of the system. The potential $U$ (7) is associated with the classical potential $e\chi$ (10) with ratio

$$\chi \stackrel{det}{=} \frac{2\alpha\beta}{\gamma}\left(\frac{1}{2\alpha\beta}\frac{|\gamma\vec{A}|^2}{2} + U + \frac{\alpha}{\beta}\frac{\Delta\sqrt{f}}{\sqrt{f}}\right). \tag{10A}$$

Note that expression (10A) contains a summand $\dfrac{\alpha}{\beta}\dfrac{\Delta\sqrt{f}}{\sqrt{f}} = -\dfrac{\hbar^2}{2m}\dfrac{\Delta\sqrt{f}}{\sqrt{f}}$ which is a well-known quantum potential in the theory of pilot wave [15-17]. Thus in (10A) we can find the term $\dfrac{|\gamma\vec{A}|^2}{2}$ corresponding to the kinetic energy of the vortex velocity field. The potentials (7) and (10A) are obtained in [14] based only on equation (1) and representation (3).

Introducing $\langle\vec{p}\rangle(\vec{r},t) \stackrel{det}{=} m\langle\vec{v}\rangle(\vec{r},t)$, then expression (10) can be formally represented in the form of the equation of Hamilton-Jacobi [26-28]

$$\frac{\hbar}{2}\frac{\partial\Phi}{\partial t} + H(\vec{r},\langle\vec{p}\rangle,t) = 0, \tag{11}$$

where

$$H(\vec{r},\langle\vec{p}\rangle,t) = \frac{|\langle\vec{p}\rangle|^2}{2m} + e\chi,$$

provided

$$H(\vec{r},\langle\vec{p}\rangle(\vec{r},t),t) = W(\vec{r},t). \tag{12}$$

Due to the view of the velocity as $\langle\vec{v}\rangle(\vec{r},t)$ (3) further instead of function $H(\vec{r},\langle\vec{p}\rangle,t)$, we will use the Hamilton function

$$\tilde{H}(\vec{r},\langle\vec{p}_p\rangle,\langle\vec{v}_s\rangle,t) = \frac{1}{2m}|\langle\vec{p}_p\rangle|^2 + (\langle\vec{p}_p\rangle,\langle\vec{v}_s\rangle) + \frac{m}{2}|\langle\vec{v}_s\rangle|^2 + e\chi(\vec{r},t), \tag{13}$$

and related with it through the Legendre transformation of the Lagrangian function $\tilde{L}(\vec{r},\langle\vec{v}\rangle,\langle\vec{v}_s\rangle,t)$:

$$\tilde{L}(\vec{r},\langle\vec{v}\rangle,\langle\vec{v}_s\rangle,t) + \tilde{H}(\vec{r},\langle\vec{p}_p\rangle,\langle\vec{v}_s\rangle,t) = (\langle\vec{v}\rangle,\langle\vec{p}_p\rangle). \tag{14}$$

From the variational analysis [26-29], we can see that action $\tilde{S} = \int \tilde{L}dt$ satisfies:

$$\frac{\partial \tilde{S}}{\partial t} + \tilde{H}\left(\vec{r}, \langle \vec{p}_p \rangle, \langle \vec{v}_s \rangle, t\right) = 0,$$

$$\nabla_r \tilde{S} = \langle \vec{p}_p \rangle, \qquad (15)$$

$$\frac{d\tilde{S}}{dt} = \tilde{L}\left(\vec{r}, \langle \vec{v} \rangle, \langle \vec{v}_s \rangle, t\right).$$

Note that due to the Helmholtz theorem (3) - (5) $\nabla_r \tilde{S} = \langle \vec{p}_p \rangle$ is correct, not $\nabla_r \tilde{S} = \langle \vec{p} \rangle$. Basing on (4), (10) for the scalar potential $\Phi$ the ratios similar to (15), are correct:

$$\frac{\hbar}{2}\frac{\partial \Phi}{\partial t} + \tilde{H}\left(\vec{r}, \langle \vec{p}_p \rangle, \langle \vec{v}_s \rangle, t\right) = 0,$$

$$\frac{\hbar}{2}\nabla_r \Phi = \langle \vec{p}_p \rangle, \qquad (16)$$

$$\frac{\hbar}{2}\frac{d}{dt}\Phi = \frac{\hbar}{2}\frac{\partial \Phi}{\partial t} + \frac{\hbar}{2}\left(\langle \vec{v} \rangle, \nabla_r \Phi\right) = -\tilde{H} + \left(\langle \vec{v} \rangle, \langle \vec{p}_p \rangle\right) = \tilde{L}\left(\vec{r}, \langle \vec{v} \rangle, \langle \vec{v}_s \rangle, t\right).$$

Comparing equations (15) and (16) we obtain:

$$\tilde{S} = \frac{\hbar}{2}\Phi + const, \qquad (17)$$

or

$$\tilde{S} = \hbar\varphi + \pi k + const,$$

that is, action $\tilde{S}$ is connected with the phase of the wave function $\varphi$. An important consequence of relation (17) is an estimation of the applicability limit of quantum and classical mechanics $\frac{\tilde{S}}{\hbar}$.

As the result on the one hand $\varphi(\vec{r},t)$ is the phase of the wave function $\Psi(\vec{r},t)$, on the other hand it is a scalar potential of the flow velocity of the probability (8), on the third hand – action (17).

§ 2 Conformal mapping $\Psi = e^Z$

Due to (2) and (9) of §1 the following is correct for functions $\Phi$ and $S$:

$$i\Phi = \operatorname{Ln}\left(\frac{\Psi}{\bar{\Psi}}\right) = \operatorname{Ln}\Psi - \operatorname{Ln}\bar{\Psi}, \qquad (1)$$

$$S = \operatorname{Ln} f = \operatorname{Ln}\left(\Psi\bar{\Psi}\right) = \operatorname{Ln}\Psi + \operatorname{Ln}\bar{\Psi}. \qquad (2)$$

We define a complex function $M$ as

$$M(\vec{r},t) \stackrel{\text{det}}{=} S(\vec{r},t) + i\Phi(\vec{r},t). \qquad (3)$$

Functions $S$ and $\Phi$ are actual functions of real variables $(\vec{r},t)$, therefore, they respectively equal to the real and imaginary parts of complex functions $M$. According to definitions (1) and (2) to functions (3) the following is correct:

$$Z \stackrel{det}{=} \frac{M}{2} = \operatorname{Ln}\Psi, \ \bar{Z} \stackrel{det}{=} \frac{\bar{M}}{2} = \operatorname{Ln}\bar{\Psi}, \qquad (4)$$

$$S = \frac{M+\bar{M}}{2}, \ i\Phi = \frac{M-\bar{M}}{2},$$

$$\Psi = e^{Z}. \qquad (5)$$

Thus, the wave function $\Psi$ is fully determined by the introduction of function $Z$. From (5) we can see that there is a set mapping in the complex plane. For mapping (5) to be univalent it is required that the domain of the function had a horizontal stripe with width of $2\pi$. By that stripe $0 < \operatorname{Im} Z < 2\pi$ appears in the complex plane slit along the ray $[0,+\infty)$ (see Fig.1). Therefore, the range of values of function $M$ will be:

$$0 < \operatorname{Im} M < 4\pi,$$
$$-\infty < \operatorname{Re} M < +\infty, \qquad (6)$$

or by definition (3)

$$0 < \Phi(\vec{r},t) < 4\pi, \qquad (7)$$
$$-\infty < S(\vec{r},t) < +\infty.$$

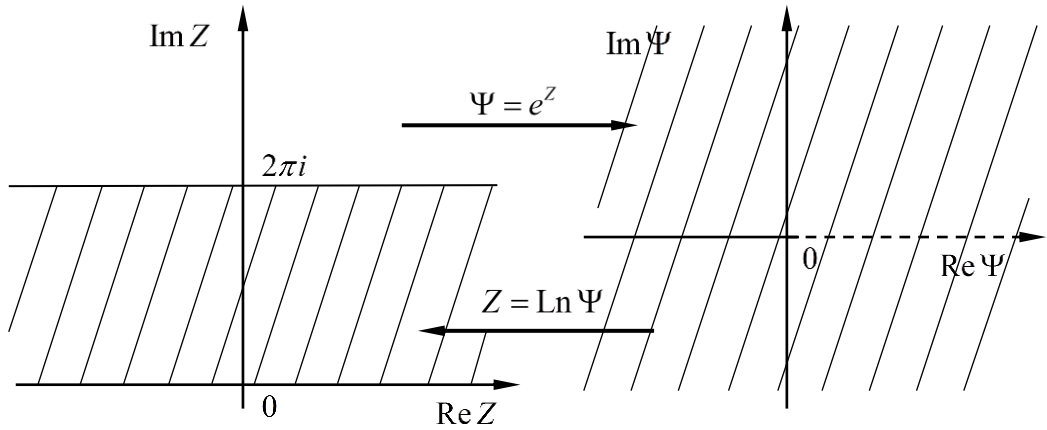

Fig. 1 The Mapping (5)

The function $\Phi(\vec{r},t)$ specifies a scalar velocity potential of the probability flow $\langle\vec{v}\rangle(\vec{r},t)$, and the function $S(\vec{r},t)$ is connected via the logarithm of a probability density function $f(\vec{r},t)$. Therefore, the function $\Psi(\vec{r},t)$ in the moment of time $t$ can be seen not only as a complex function of a real variable $\vec{r}$ but as a complex function of a complex variable $M = 2Z$ with domain of definition (6)-(7), that is $\Psi(M)$.

The range of values of the scalar potential $\Phi(\vec{r},t)$ in the general case can be wider than (7), in this case we can see violation of univalent mapping (5). The inverse mapping is a function

Ln Ψ that is defined on Riemann surface (see Fig.2). The function $\Phi(\vec{r},t)$ has the physical meaning of the action, the phase of the wave function and the velocity potential of the flow probabilities. Let us consider $\Phi(\vec{r},t)$ as an action, then the limiting transition between classical and quantum mechanics, defined by relation $\frac{S}{\hbar}=\frac{\Phi}{2}=\varphi$ (17) §1 will be associated with the domain of univalent wave function Ψ (Ln Ψ).

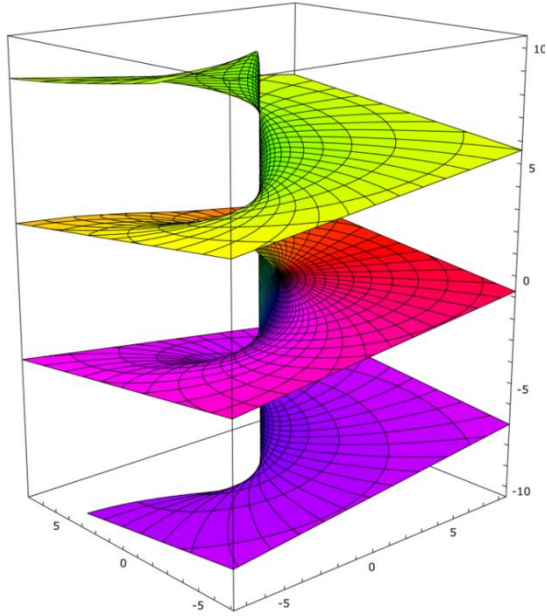

Fig. 2 Riemann surfaces for function $Z = \text{Ln}\,\Psi$

In classical mechanics $\varphi \gg 1$, that is, the function Ψ (Ln Ψ) is multivalent, as the domain of the function Ψ becomes wider than the horizontal stripe $2\pi$ (see Fig.1,2). The value of the function Ψ will take several leaves of the Riemann surface.

In quantum mechanics $\varphi \sim 1$, that is, the function Ψ (Ln Ψ) is univalent (see Fig.1,2), and $\varphi$ does not exceed $2\pi < 10$. If a violation of univalence, for example, in the case of the two-valence function $0 < \varphi < 4\pi$. Value $4\pi > 10 \gg 1$ that is the «movement» in the direction of classical mechanics.

As $\frac{\partial \Psi(M)}{\partial M} = \frac{1}{2}e^{\frac{M}{2}} \neq 0$ for all $M$ from the domain of definition (6), (7), then function $\Psi(M)$ sets the conformal mapping, i.e. preserve angles between curves and has the property of constancy of strains in the neighborhood of point $M$. Consequently, the Jacobian of such mapping shall be different from zero, that is

$$\Psi = e^{\frac{S}{2}} e^{i\frac{\Phi}{2}} = u(S,\Phi) + iv(S,\Phi),$$

$$u(S,\Phi) = e^{\frac{S}{2}}\cos\left(\frac{\Phi}{2}\right),\quad v(S,\Phi) = e^{\frac{S}{2}}\sin\left(\frac{\Phi}{2}\right),$$

$$J = \begin{vmatrix} \frac{\partial u}{\partial S} & \frac{\partial u}{\partial \Phi} \\ \frac{\partial v}{\partial S} & \frac{\partial v}{\partial \Phi} \end{vmatrix} = \begin{vmatrix} \frac{1}{2}e^{\frac{S}{2}}\cos\left(\frac{\Phi}{2}\right) & -\frac{1}{2}e^{\frac{S}{2}}\sin\left(\frac{\Phi}{2}\right) \\ \frac{1}{2}e^{\frac{S}{2}}\sin\left(\frac{\Phi}{2}\right) & \frac{1}{2}e^{\frac{S}{2}}\cos\left(\frac{\Phi}{2}\right) \end{vmatrix} = \qquad (8)$$

$$= \frac{1}{4}e^{S}\cos^{2}\left(\frac{\Phi}{2}\right) + \frac{1}{4}e^{S}\sin^{2}\left(\frac{\Phi}{2}\right) = \frac{e^{S}}{4} = \frac{1}{4}f = \frac{1}{4}|\Psi|^{2} \neq 0,$$

or

$$J = \left|\frac{\partial \Psi(M)}{\partial M}\right|^{2} = \frac{1}{4}\left|e^{\frac{S}{2}}\right|^{2}\left|e^{i\frac{\Phi}{2}}\right|^{2} = \frac{e^{S}}{4} = \frac{1}{4}|\Psi|^{2} \neq 0.$$

From (8) it follows that the Jacobian $J$ of mapping $\Psi(M) = e^{M/2}$ (5) is actually a probability density $|\Psi|^{2}$, that is,

$$|\Psi|^2 = 4J. \tag{9}$$

As the probability density $f = |\Psi|^2$ satisfies continuity equation (1) §1, then from (9) it follows that the same statement holds for the Jacobian $J$

$$\frac{\partial J}{\partial t} + \operatorname{div}_r \left[ J \langle \vec{v} \rangle \right] = 0. \tag{10}$$

As the Jacobian specifies the coefficient of proportionality between $dudv$ and $dSd\Phi$ at mapping $\Psi$, then from (9) it follows that the probability density $|\Psi|^2$ determines the aspect ratio domains in mapping (5). Here the following ratios are correct

$$\omega(D') = \iint_{D'} dudv = \iint_D |J| dSd\Phi = \frac{1}{4} \iint_D |\Psi|^2 dSd\Phi = \frac{1}{4} \iint_D e^S dSd\Phi. \tag{11}$$

Selecting «half» the field of definition (6) for $D$, that is, $-\infty < \operatorname{Re} M < 0$, then domain $D'$ is a unit circle with a cut along the segment $[0,1]$, and integral (11) is equal to $\pi$, which corresponds to the domain of a circle with single radius.

### § 3 Complex principle of least action (CPLA)

The value of the function $\Psi$ at each point in the moment of time $t_1$ depends on physical coordinate $\vec{r}$. For each point of the continuum with coordinate $\vec{r}$ there is a value of density $f(\vec{r}, t_1)$ and velocity potential $\Phi(\vec{r}, t_1)$, which determine the value of the function $Z(\vec{r}, t_1)$ by the formula (3)-(4) §2. In this case physical system will correspond to some domain $D_{Z_1}$ of values of function $Z(\vec{r}, t_1)$. Conformal mapping (5) §2 will transform domain $D_{Z_1}$ into $D_{\Psi_1}$ of values of function $\Psi(\vec{r}, t_1)$ (or by (5) §2 $\Psi(Z(\vec{r}, t_1)) = \Psi(Z_1)$).

In the next moment of time $t_2 > t_1$ coordinate $\vec{r}$ will change the value of the density $f(\vec{r}, t_2)$ and the potential $\Phi(\vec{r}, t_2)$, therefore, the range of values of function $Z$ goes from $D_{Z_1}$ to $D_{Z_2}$. A similar change will occur with the range of values of function $\Psi$, that is, domain $D_{\Psi_1}$ will go to domain $D_{\Psi_2}$ (see Fig. 3).

With such a mapping in terms of Lagrangian coordinates, every point $\xi_1 \in D_{\Psi_1}$ will move to some point $\xi_2 \in D_{\Psi_2}$. Let us introduce a continuous parameter $t_1 \leq \tau \leq t_2$. Let us build the mapping of $D_Z$ and $D_\Psi$ for each value of the parameter $\tau$ (assuming that is possible). Then every point $\xi_1 \in D_{\Psi_1}$ will move to the point $\xi_2 \in D_{\Psi_2}$ according to some continuous paths $\xi(\tau)$ lying in the complex plane.

$$\xi = \xi(\tau), \quad \xi(t_1) = \xi_1, \quad \xi(t_2) = \xi_2, \quad t_1 \leq \tau \leq t_2. \tag{1}$$

Each point $D_{\Psi_1}$ moves along its path $\xi(\tau)$ to the point domain $D_{\Psi_2}$. Thus, when mapping (5) §2 there is an infinite number of trajectories $\xi(\tau)$ along which domain $D_{\Psi_1}$ goes to $D_{\Psi_2}$ (see Fig. 3).

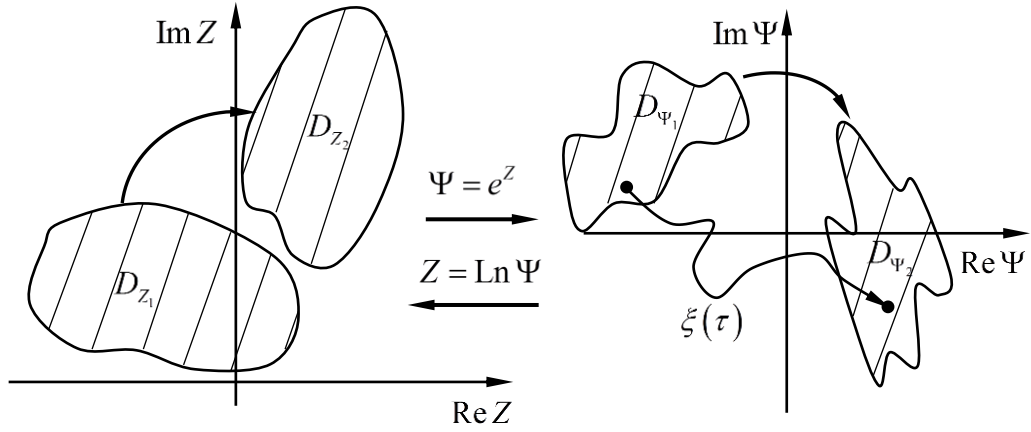

Fig. 3 Mapping $D_{Z_1} \to D_{Z_2}$ and $D_{\Psi_1} \to D_{\Psi_2}$

This mapping is associated with a change of the distribution function of the density $f(\vec{r},t)$, which can be interpreted in two ways. On the one hand it is the density of mass, charge, and on the other hand it is the probability density of a single «particle» («quantum particles»).

Due to the mentioned above about the trajectories $\xi(\tau)$ along which the points $D_\Psi$ move around it turns out that a quantum particle does not have any *single* trajectory, as a quantum particle is approximated by a distribution $f(\vec{r},t)$ with this movement at the same time there is an infinite number of trajectories $\xi(\tau)$ along which domain $D_{\Psi_1}$ goes to $D_{\Psi_2}$. Obviously, we can find the average trajectory $\langle \xi(\tau) \rangle$ corresponding to the motion of the «center of mass» (the most probable trajectory) of such distribution. From this point of view typical sizes with which quantum mechanics deals are comparable to the standard deviations $\sigma$ for the probability density function $f(\vec{r},t)$. In this case, the size of the object cannot be considered as a point like in classical mechanics, as its characteristic size is comparable to $\sigma$. Therefore, the notion of such object trajectory is undefined.

In the complex analysis theory for function $\mathrm{Ln}$, there is a representation in the form of an integral along a path in the complex plane. Function $Z$ in accordance with (5) §2 is expressed using $\Psi$ as $Z = \mathrm{Ln}\,\Psi$, therefore, the following is correct

$$Z = \mathrm{Ln}\,\Psi = \int_1^\Psi \frac{d\xi}{\xi}. \qquad (2)$$

The expanded function in (2) has a pole of first order at zero, so depending on the path of integration the value of integral (2) can be different (see Fig. 4).

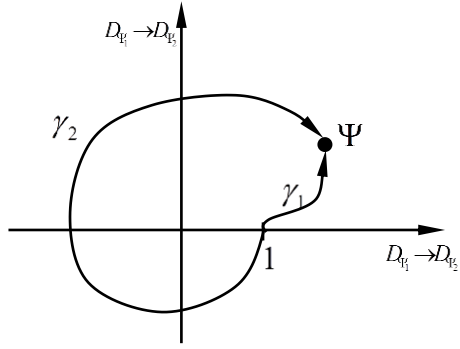

Fig. 4 The path of integration (2)

Let contour to be $\gamma = \gamma_1 \cup \bar{\gamma}_2$, where $\bar{\gamma}_2$ is the $\gamma_2$ path with the opposite orientation, then for expression (2) in accordance with Fig.4 we obtain

$$\oint_\gamma \frac{d\xi}{\xi} = \int_{\gamma_1} \frac{d\xi}{\xi} - \int_{\gamma_2} \frac{d\xi}{\xi} = 2\pi i \operatorname{res}\left[\frac{1}{\xi}, 0\right] = 2\pi i,$$

$$\int_{\gamma_1} \frac{d\xi}{\xi} = 2\pi i + \int_{\gamma_2} \frac{d\xi}{\xi}, \tag{3}$$

that is, path integrals $\gamma_1$ and $\gamma_2$ differ by $2\pi i$. If the contour $\gamma$ does not contain the origin (the function pole $1/\xi$) then the path integrals $\gamma_1$ and $\gamma_2$ coincide. Expression (3) implies that the real part of the path integrals $\gamma_1$ and $\gamma_2$ coincide, and the imaginary parts differ by $2\pi$. This difference is connected with the fact that $\operatorname{Im} Z = \Phi$, $\Phi$ as the phase is determined with accuracy up to $2\pi k$.

Let us consider the motion along one of the trajectories $\xi(\tau)$ connecting $\Psi_1$ and $\Psi_2$. In moments of time $t_1$ and $t_2$ for function $Z$ in accordance with (2) we can have the following (see Fig.5)

$$Z_1 = \operatorname{Ln} \Psi_1 = \int_1^{\Psi_1} \frac{d\xi}{\xi}, \quad Z_2 = \operatorname{Ln} \Psi_2 = \int_1^{\Psi_2} \frac{d\xi}{\xi},$$

$$Z_{12} \stackrel{\text{det}}{=} Z_2 - Z_1 = \operatorname{Ln} \Psi_2 - \operatorname{Ln} \Psi_1 = \operatorname{Ln}\left(\frac{\Psi_2}{\Psi_1}\right) = \operatorname{Ln} \Psi_{12}, \tag{4}$$

where $\Psi_{12} \stackrel{\text{det}}{=} \frac{\Psi_2}{\Psi_1}$. Due to (3) and (4) §2 the following is correct

$$Z_{12} = \int_1^{\Psi_2} \frac{d\xi}{\xi} - \int_1^{\Psi_1} \frac{d\xi}{\xi} = \int_{\Psi_1}^{\Psi_2} \frac{d\xi}{\xi}, \tag{5}$$

$$Z_{12} = Z_2 - Z_1 = \frac{1}{2}(S_2 + i\Phi_2) - \frac{1}{2}(S_1 + i\Phi_1) = \frac{S_2 - S_1}{2} + i\frac{\Phi_2 - \Phi_1}{2} =$$

$$= \frac{\operatorname{Ln}|\Psi_2|^2 - \operatorname{Ln}|\Psi_1|^2}{2} + \frac{1}{2}\operatorname{Ln}\left(\frac{\Psi_2}{\overline{\Psi}_2}\right) - \frac{1}{2}\operatorname{Ln}\left(\frac{\Psi_2}{\overline{\Psi}_2}\right) = \frac{1}{2}\operatorname{Ln}\left|\frac{\Psi_2}{\Psi_1}\right|^2 + \frac{1}{2}\operatorname{Ln}\left(\frac{\Psi_2}{\overline{\Psi}_2}\frac{\overline{\Psi}_1}{\Psi_1}\right) =$$

$$= \frac{1}{2}\left[\operatorname{Ln}|\Psi_{12}|^2 + \operatorname{Ln}\left(\frac{\Psi_{12}}{\overline{\Psi}_{12}}\right)\right] = \frac{S_{12} + i\Phi_{12}}{2},$$

$$Z_{12} = \frac{1}{2}(S_{12} + i\Phi_{12}), \tag{6}$$

where $S_{12} \stackrel{\text{det}}{=} \operatorname{Ln}|\Psi_{12}|$, $i\Phi_{12} \stackrel{\text{det}}{=} \operatorname{Ln}\left(\frac{\Psi_{12}}{\overline{\Psi}_{12}}\right)$.

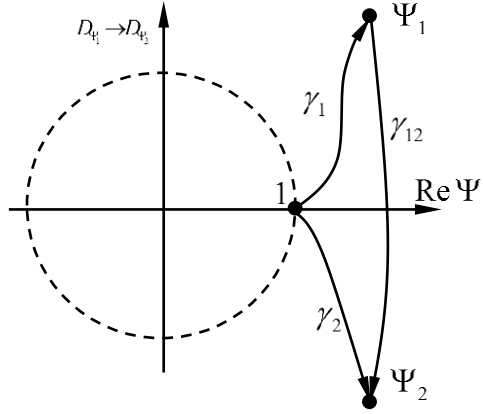

Fig. 5 The path of integration (4)

From (2) and (4) it follows that:

$$Z_{12} = \int_1^{\Psi_{12}} \frac{d\xi}{\xi}. \tag{7}$$

Let us transform expression (7) using parameterization (1)

$$Z_{12} = \frac{S_{12}}{2} + i\frac{\Phi_{12}}{2} = \int_{t_1}^{t_2} \frac{\xi'}{\xi} d\tau = \int_{t_1}^{t_2} (\operatorname{Ln}\xi)' d\tau = \int_{t_1}^{t_2} \operatorname{Re}(\operatorname{Ln}\xi)' d\tau + i\int_{t_1}^{t_2} \operatorname{Im}(\operatorname{Ln}\xi)' d\tau, \tag{8}$$

that is

$$\frac{S_{12}}{2} = \int_{t_1}^{t_2} \operatorname{Re}(\operatorname{Ln}\xi)' d\tau, \quad \frac{\Phi_{12}}{2} = \int_{t_1}^{t_2} \operatorname{Im}(\operatorname{Ln}\xi)' d\tau. \tag{9}$$

Let us transform the integrals (9). For the first integral from (9) we obtain

$$\frac{S_{12}}{2} = \int_{t_1}^{t_2} \operatorname{Re}(\operatorname{Ln}\xi)' d\tau = \int_{t_1}^{t_2} (\ln|\xi|)' d\tau = \ln|\xi|\Big|_1^{\Psi_{12}} = \ln|\Psi_{12}|, \tag{10}$$

from here

$$S_{12} = 2\ln|\Psi_{12}| = \text{Ln}|\Psi_{12}|^2,$$

which corresponds to definition $S_{12}$ (6). For $S_{12}$ the following integral representation is correct

$$\frac{S_{12}}{2} = \frac{S_2}{2} - \frac{S_1}{2} = \frac{1}{2}\int_{t_1}^{t_2}\left(\frac{dS}{d\tau}\right)d\tau = -\frac{1}{2}\int_{t_1}^{t_2} Q d\tau$$

$$S_{12} = -\int_{t_1}^{t_2} Q d\tau, \qquad (11)$$

which corresponds to the solution of equation (2) §1.

The second integral of (8) is associated with the principle of least action (PLA), as function $\Phi_{12}$ is an action $\tilde{S}_{12}$ in accordance with (17) §1. The result:

$$\frac{\Phi_{12}}{2} = \int_{t_1}^{t_2}\text{Im}(\text{Ln}\,\xi)' d\tau = \frac{\tilde{S}_{12}}{\hbar} = \frac{1}{\hbar}\int_{t_1}^{t_2}\tilde{L}d\tau, \qquad (12)$$

hence, on the trajectory

$$\text{Im}(\text{Ln}\,\xi)' = \frac{1}{\hbar}\tilde{L}. \qquad (13)$$

Expressions (12) and (13) can be further transformed into

$$\text{Im}(\text{Ln}\,\xi)' = \text{Arg}\,\xi' = \frac{1}{\hbar}\tilde{L} = \frac{d}{dt}\frac{\Phi_{12}}{2}$$

$$\frac{\Phi_{12}}{2} = \int_{t_1}^{t_2}\text{Im}(\text{Ln}\,\xi)' d\tau = \int_{t_1}^{t_2}\text{Arg}\,\xi' d\tau = \text{Arg}\,\xi\big|_1^{\Psi_{12}} = \text{Arg}\,\Psi_{12} - \text{Arg}\,1 = \text{Arg}\,\Psi_{12} = \varphi_{12},$$

that corresponds the definition of function $\Phi_{12}$ as the double phase of the wave function $\Psi_{12}$ (8) §1. It turns out that the Lagrangian can be represented as

$$\tilde{L} = \hbar\,\text{Im}(\text{Ln}\,\Psi_{12})' = \hbar\,\text{Im}(Z_{12})', \qquad (14)$$

or

$$\tilde{L} = \hbar\,\text{Im}(Z_{12})' = \frac{\hbar}{2}\text{Im}(S_{12} + i\Phi_{12})' = \frac{\hbar}{2}\frac{d}{dt}\Phi_{12},$$

that also true for $\Phi_{12}$, if it is considered as the action (16) §1.

Let us return to integral (8), taking (11) and (12) into account we obtain

$$Z_{12} = -\frac{1}{2}\int_{t_1}^{t_2} Q d\tau + i\frac{1}{\hbar}\int_{t_1}^{t_2}\tilde{L}d\tau, \qquad (15)$$

or

$$Z_{12} = \frac{S_{12}}{2} + i\frac{\tilde{S}_{12}}{\hbar} = \ln|\Psi_{12}| + i\frac{\tilde{S}_{12}}{\hbar}. \tag{16}$$

Expression (16) in accordance with mapping (5) §2 gives an idea of the wave function $\Psi_{12}$ in the form of

$$\Psi_{12} = e^{Z_{12}} = |\Psi_{12}| e^{i\frac{\tilde{S}_{12}}{\hbar}}. \tag{17}$$

Expression (15) can be formally interpreted as a complex action, the imaginary part of which is a classical action, and requires to be studied. If value $Q = 0$, then expression (15) will go to a classical action, and the equation for the probability density (1) §1 will be in the transfer equation (2A) §1.

It was shown above that the action $Z_{12}$ integral (7) does not depend on the integration contour (with the accuracy up to $2\pi i$), hence it is a constant complex value. Variation of a constant value equals to zero, that is $\delta Z_{12} \equiv 0$.

Thus, it is possible to formulate the complex principle of least action (CPLA). Note that it is necessary to distinguish the concept of the integral along the trajectory used in the classical principle of least action (PLA), and the notion of integral along the trajectory in the complex plane (15).

Let us formally apply PLA to the action (15). Action (15) has a complex value, therefore:

$$\delta Z_{12} = 0 \Leftrightarrow \begin{cases} \delta(\operatorname{Re} Z_{12}) = 0, \\ \delta(\operatorname{Im} Z_{12}) = 0. \end{cases} \tag{18}$$

Condition $\delta(\operatorname{Im} Z_{12}) = 0$ is equivalent to the classical PLA ($\delta \tilde{S}_{12} = 0$) and leads to the Euler-Lagrange equation for function $\tilde{L}$.

Let us consider condition $\delta(\operatorname{Re} Z_{12}) = 0$. For expanded function $Q$ the following is correct (2)§1. Taking into account that the velocity $\langle \vec{v} \rangle$ admits decomposition (3), (4) §1, the expression for $Q$ takes the form:

$$Q = \operatorname{div}_r \langle \vec{v} \rangle = -\alpha \Delta_r \Phi_{12} + \gamma \operatorname{div}_r \vec{A}. \tag{19}$$

Next, we consider two types of gauges (Lorenz and Coulomb)

$$\frac{\varepsilon \mu}{c^2} \frac{\partial \chi}{\partial t} + \operatorname{div}_r \vec{A} = 0, \tag{20}$$

and

$$\operatorname{div}_r \vec{A} = 0. \tag{21}$$

From the point of view of the field theory, the Lorenz gauge (20), unlike (21) was fulfilled in all inertial reference frames. If the Lorenz gauge was carried out (20), then expression (19) takes the form

$$Q = \frac{\hbar}{2m} \Delta_r \Phi_{12} + e \frac{\varepsilon \mu}{mc^2} \frac{\partial \chi}{\partial t} = \frac{1}{m} \Delta_r \tilde{S}_{12} + e \frac{\varepsilon \mu}{mc^2} \frac{\partial \chi}{\partial t}. \tag{22}$$

Calculating $\delta(\operatorname{Re} Z_{12}) = 0$ accounting (22) and conditions $\delta \tilde{S}_{12} = 0$, we obtain

$$\delta(\operatorname{Re} Z_{12}) = -\frac{1}{2}\int_{t_1}^{t_2} \delta Q d\tau = -\frac{1}{2}\int_{t_1}^{t_2} \delta\left[\frac{1}{m}\Delta_r \tilde{S}_{12} + e\frac{\varepsilon\mu}{mc^2}\frac{\partial \chi}{\partial t}\right]d\tau = \qquad (23)$$

$$= -\frac{1}{2m}\int_{t_1}^{t_2} \Delta_r\left(\delta \tilde{S}_{12}\right)d\tau - e\frac{\varepsilon\mu}{2mc^2}\int_{t_1}^{t_2}\frac{\partial}{\partial t}\delta\chi d\tau = -\frac{\varepsilon\mu}{2mc^2}\int_{t_1}^{t_2}\frac{\partial}{\partial t}(\nabla_r e\chi)\delta\vec{r}d\tau = 0.$$

that is

$$-\frac{\partial}{\partial t}(\nabla_r e\chi) = 0 \Rightarrow -\nabla_r(e\chi) = \vec{F}(\vec{r}), \qquad (24)$$

where $e\chi$ – the value of the potential energy of the classical (non-quantum) system (10A) §1, therefore, value $\vec{F}(\vec{r})$ in (24) corresponds to the potential force, which does not depend on time explicitly.

Thus, the complex principle of least action (CPLA) using the Lorenz gauge (20) leads to the classical Euler-Lagrange equation for the trajectory and to the limit of the potential force type (24).

Expression (22) for Coulomb gauge (21) takes the form

$$Q = -\alpha\Delta_r \Phi_{12} = \frac{\hbar}{2m}\Delta_r\Phi_{12} = \frac{1}{m}\Delta_r\tilde{S}_{12}. \qquad (25)$$

Substituting expression for $Q$ (25) into condition $\delta(\operatorname{Re} Z_{12}) = 0$ and considering $\delta\tilde{S}_{12} = 0$, we obtain the correct identity, without limitations on the potential force type.

As the result CPLA for both gauges (20) and (21) gives the Euler-Lagrangian equation, but with different restrictions on potential forces.

Note that as a complex action (15) the wave function defined by mapping (5) §2 can be considered. In this case, the search of the minimum of the wave function of ($\delta\Psi_{12} = 0$) for all complex trajectories will be reduced to the minimum search $Z_{12}$ ($\delta Z_{12} = 0$), as

$$\delta\Psi_{12} = \delta e^{Z_{12}} = e^{Z_{12}}\delta Z_{12} = \Psi_{12}\delta Z_{12} = 0 \Leftrightarrow \delta Z_{12} = 0. \qquad (26)$$

Complex action $Z_{12}$ can be written in another form based on direct definitions of functions $S_{12}$ and $\Phi_{12}$, for example (see Fig. 6)

$$Z_{12} = \frac{1}{2}(S_{12} + i\Phi_{12}) = \frac{1}{2}\operatorname{Ln}(\Psi_{12}\bar{\Psi}_{12}) + \frac{1}{2}\operatorname{Ln}\left(\frac{\Psi_{12}}{\bar{\Psi}_{12}}\right) = \frac{1}{2}(\operatorname{Ln}\bar{\Psi}_{12} + \operatorname{Ln}\Psi_{12}) +$$

$$+\frac{1}{2}(\operatorname{Ln}\Psi_{12} - \operatorname{Ln}\bar{\Psi}_{12}) = \frac{1}{2}\left(\operatorname{Ln}\bar{\Psi}_{12} - \operatorname{Ln}\frac{1}{\Psi_{12}}\right) + \frac{1}{2}(\operatorname{Ln}\Psi_{12} - \operatorname{Ln}\bar{\Psi}_{12}) =$$

$$= \frac{1}{2}\operatorname{Ln}\xi\Big|_{1/\Psi_{12}}^{\bar{\Psi}_{12}} + \frac{1}{2}\operatorname{Ln}\xi\Big|_{\bar{\Psi}_{12}}^{\Psi_{12}} = \frac{1}{2}\int_{1/\Psi_{12}}^{\bar{\Psi}_{12}}\frac{d\xi}{\xi} + \frac{1}{2}\int_{\bar{\Psi}_{12}}^{\Psi_{12}}\frac{d\xi}{\xi},$$

$$Z_{12} = \frac{1}{2} \int_{1/\Psi_{12}}^{\Psi_{12}} \frac{d\xi}{\xi} = \frac{1}{2} \int_{1/\Psi_{12}}^{\bar{\Psi}_{12}} \frac{d\xi}{\xi} + \frac{1}{2} \int_{\bar{\Psi}_{12}}^{\Psi_{12}} \frac{d\xi}{\xi}. \qquad (27)$$

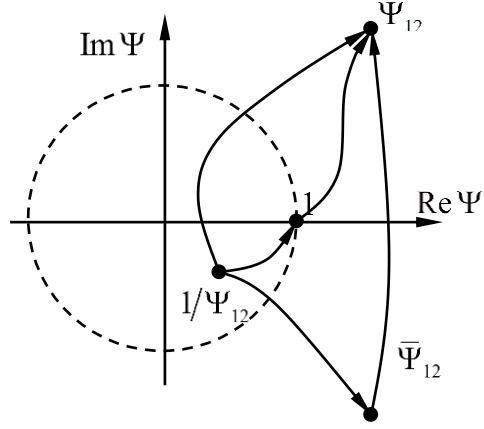

Fig. 6 The path of integration (27)

From comparison of the integrals (7) and (27) it follows that (see Fig.6)

$$Z_{12} = \frac{1}{2} \int_{1/\Psi_{12}}^{\Psi_{12}} \frac{d\xi}{\xi} = \frac{1}{2} \int_{1/\Psi_{12}}^{1} \frac{d\xi}{\xi} + \frac{1}{2} \int_{1}^{\Psi_{12}} \frac{d\xi}{\xi} = \frac{1}{2} \int_{1/\Psi_{12}}^{1} \frac{d\xi}{\xi} + \frac{1}{2} Z_{12},$$

$$\frac{1}{2} \int_{1/\Psi_{12}}^{1} \frac{d\xi}{\xi} = Z_{12} - \frac{1}{2} Z_{12} = \frac{1}{2} Z_{12},$$

$$Z_{12} = \int_{1/\Psi_{12}}^{1} \frac{d\xi}{\xi} = \int_{1}^{\Psi_{12}} \frac{d\xi}{\xi}. \qquad (28)$$

The value of integral $Z_{12}$ does not depend on the path of integration, but depends on the position of the ends of the trajectory. If $|\Psi_{12}| = 1$ then the points 1, $\Psi_{12}$, $1/\Psi_{12}$ are located on the unit circle, and their arguments differ by $\pi/4$. Action (28) in this case is equal to $i\frac{\pi}{4}$, i.e. is purely imaginary, therefore, $S_{12} = 0 = \ln \frac{f(\vec{r},t_1)}{f(\vec{r},t_2)} = \ln |\Psi_{12}|^2$ that corresponds to $\frac{f(\vec{r},t_1)}{f(\vec{r},t_2)} = 1$, that is

$$f(\vec{r},t_1) = f(\vec{r},t_2). \qquad (29)$$

Equation (29) indicates the stationary state of the system at the point $\vec{r}$ as the obtained results correspond to one of the infinite number of trajectories. The wave function $\Psi_{12}$ corresponds to one of the trajectories $\xi(\tau)$ and is associated with the probability of transition from one «point» of the system from state «1» to «2». Considering all the trajectories $\xi(\tau)$, along which domain $D_{\Psi_1}$ goes to domain $D_{\Psi_2}$, we can then define a generalized complex action $\tilde{Z}_{12}$ that contains the integrals over all trajectories $\xi(\tau)$.

Let us consider functional set X, which includes all the trajectories $\xi(\tau)$, along which domain $D_{\Psi_1}$ goes to domain $D_{\Psi_2}$. In the general case, the cardinality of the set X is a continuum, that is $|X| = \mathfrak{c}$. The functionality $Z_{12} = Z_{12}[\xi]$ (16) is defined on the functional set X. That is, for each $\xi(\tau) \in X$ by the formula (16) $Z_{12}[\xi]$ can be calculated. Due to (17) functionality $\Psi_{12} = \Psi_{12}[\xi] = e^{Z_{12}[\xi]}$ is also defined on a set X. For functionality $\Psi_{12}[\xi]$ the following relation is valid (17)

$$\Psi_{12}[\xi] = |\Psi_{12}[\xi]| e^{i\frac{\tilde{S}_{12}[\xi]}{\hbar}}, \tag{30}$$

$$|\Psi_{12}[\xi]| = \frac{|\Psi_2|}{|\Psi_1|}, \tag{31}$$

where $\Psi_1$ is the beginning and $\Psi_2$ – the end of trajectory $\xi(\tau)$. For physical trajectory $\vec{r}(t)$, this means that

$$|\Psi_1|^2 = f(\vec{r}(t_1), t_1), \; |\Psi_2|^2 = f(\vec{r}(t_2), t_2). \tag{32}$$

If the probability density is conserved $f(\vec{r}(t_1), t_1) = f(\vec{r}(t_2), t_2)$ along the trajectories $\vec{r}(t)$ (characteristics), for example, the transport equation (2A) §1 for function $f(\vec{r}, t)$, then expression (31) takes the form

$$|\Psi_{12}[\xi]| = \frac{|\Psi_2|}{|\Psi_1|} = \frac{f(\vec{r}(t_2), t_2)}{f(\vec{r}(t_1), t_1)} = 1. \tag{33}$$

Substituting (33) into (30) for $\Psi_{12}[\xi]$ functionality, we obtain

$$\Psi_{12}[\xi] = e^{i\frac{\tilde{S}_{12}[\xi]}{\hbar}}. \tag{34}$$

From (34) it follows that the motion along path $\xi(\tau)$ from values $\Psi_1$ to value $\Psi_2$ provided (33) is equivalent to the rotation in the complex plane of the number $\Psi_1$ on the angle $\Delta\varphi = \varphi_2 - \varphi_1$, where $\varphi_i = \arg \Psi_i$, $i = 1, 2$.

Using (26) on the equivalence of the functionality minimum search $\Psi_{12}[\xi]$ and $Z_{12}[\xi]$ we shall consider only functionality (34) defined on the set X. To account for the contribution from all trajectories (for all the elements (points)) of the functional set X we sum up contributions (34) over the entire set X. As the cardinality of the set $|X| = \mathfrak{c}$ (continuum), it would be logical to take the continual integral from (17), (30) as the sum

$$\tilde{\Psi}_{12} \stackrel{\text{det}}{=} \int_1^2 e^{i\frac{Z_{12}[\xi]}{\hbar}} D\xi = \int_1^2 |\Psi_{12}(\xi)| e^{i\frac{\tilde{S}_{12}[\xi]}{\hbar}} D\xi, \tag{35}$$

or accounting conditions (33)

$$\tilde{\Psi}_{12} = \int_1^2 e^{i\frac{\tilde{S}_{12}[\xi]}{\hbar}} D\xi, \qquad (36)$$

where $\int_1^2 D\xi$ − conditional note of the infinitely divisible functional integration over all trajectories $\xi(\tau) \in X$ transforming the $D_{\Psi_1}$ into the $D_{\Psi_2}$.

Integral (36) in fact, is known as the Feynman's path integral. Note that integral (36) is obtained as a special case of integral (35) taking the conservation of probability density along the trajectories (characteristics) into account.

## § 4 The evolution operator

Let us consider the mapping of $D_{\Psi_1}$ to $D_{\Psi_2}$ and related ratios (30), (31) §3. Let us calculate value $\Psi_{12}[\xi]\Psi_1$ using ratio (4), (6), (12), (31)§3 and we obtain

$$\Phi_{12} = \text{Arg}\left(\frac{\Psi_{12}}{\bar{\Psi}_{12}}\right) = \text{Arg}\left(\frac{\Psi_2}{\Psi_1}\frac{\bar{\Psi}_1}{\bar{\Psi}_2}\right) = \text{Arg}\left(\frac{|\Psi_2|e^{i\varphi_2}}{|\Psi_1|e^{i\varphi_1}}\frac{|\bar{\Psi}_1|e^{-i\varphi_1}}{|\bar{\Psi}_2|e^{-i\varphi_2}}\right) =$$
$$= \text{Arg}\left(e^{i2(\varphi_2-\varphi_1)}\right) = 2(\varphi_2-\varphi_1) + 2\pi k = \Phi_2 - \Phi_1, \qquad (1)$$

considering (1) and $e^{i2\pi k} = 1$ we obtain

$$\Psi_{12}[\xi]\Psi_1 = |\Psi_{12}[\xi]|e^{i\frac{\tilde{S}_{12}[\xi]}{\hbar}}\Psi_1 = \frac{|\Psi_2|}{|\Psi_1|}e^{i\frac{\Phi_{12}}{2}}|\Psi_1|e^{i\frac{\Phi_1}{2}} = |\Psi_2|e^{i(\varphi_2-\varphi_1)}e^{i\varphi_1} = |\Psi_2|e^{i\varphi_2},$$
$$\Psi_{12}[\xi]\Psi_1 = \Psi_2. \qquad (2)$$

From (2) it is seen that $\Psi_{12}[\xi]$ transforms $\Psi_1$ from the domain of the initial state $D_{\Psi_1}$ into the endpoint $\Psi_2$ domain of the state $D_{\Psi_2}$. Thus, $\Psi_{12}[\xi]$ can be interpreted as the mapping operator of the initial state domain $D_{\Psi_1}$ into the domain of the endpoint condition $D_{\Psi_2}$. Let us convert the view of operator $\Psi_{12}[\xi]$. Starting with action $\tilde{S}_{12}$. Using the Legendre transformation (14) §1 we obtain

$$\tilde{L} = -\tilde{H} + (\langle\vec{v}\rangle, \langle\vec{p}_p\rangle),$$
$$\tilde{S}_{12} = \int_{t_1}^{t_2}\tilde{L}d\tau = -\int_{t_1}^{t_2}\tilde{H}d\tau + \int_{t_1}^{t_2}(\langle\vec{v}\rangle, \langle\vec{p}_p\rangle)d\tau = -\int_{t_1}^{t_2}\tilde{H}d\tau + \frac{\hbar}{2}\int_{r_1}^{r_2}(\nabla\Phi, d\vec{r}) = \qquad (3)$$
$$= -\int_{t_1}^{t_2}\tilde{H}d\tau + \int_{r_1}^{r_2}(\langle\vec{p}_p\rangle, d\vec{r}).$$

Substituting (3) into (34) §3, and introducing the notation

$$\mathrm{T} = \left|\Psi_{12}[\xi]\right| e^{\frac{i}{\hbar}\int_{t_1}^{t_2}(\langle \vec{p}_p \rangle, d\vec{r})} \overset{\det}{.} \quad (4)$$

For $\Psi_{12}[\xi]$ we obtain the following

$$\Psi_{12}[\xi] = \left|\Psi_{12}[\xi]\right| e^{i\frac{\tilde{S}_{12}[\xi]}{\hbar}} = \left|\Psi_{12}[\xi]\right| e^{\frac{i}{\hbar}\int_{t_1}^{t_2}(\langle \vec{p}_p \rangle, d\vec{r})} e^{-\frac{i}{\hbar}\int_{t_1}^{t_2} \tilde{H} d\tau},$$

$$\Psi_{12}[\xi] = \mathrm{T} e^{-\frac{i}{\hbar}\int_{t_1}^{t_2} \tilde{H} d\tau}. \quad (5)$$

Expression (5) formally looks like the evolution operator $\hat{S}(t_2, t_1)$ in quantum mechanics

$$\hat{S}(t_2, t_1) = \hat{T} e^{-\frac{i}{\hbar}\int_{t_1}^{t_2} \tilde{H} d\tau}, \quad t_2 > t_1, \quad (6)$$

$$\hat{S}(t_2, t_1) = \overline{\hat{T}} e^{\frac{i}{\hbar}\int_{t_1}^{t_2} \tilde{H} d\tau}, \quad t_2 < t_1,$$

where $\hat{T}, \overline{\hat{T}}$ – the operators of ordering and anti-ordering respectively. As well as $\Psi_{12}[\xi]$ (2) operator $\hat{S}(t_2, t_1)$ (6) produces the transfer from state «1» to «2»

$$\hat{S}(t_2, t_1)\left|\Psi(t_1)\right\rangle = \left|\Psi(t_2)\right\rangle. \quad (7)$$

## § 5 The Bohr-Sommerfeld quantization rule

Let us consider the Bohr-Sommerfeld quantization principle [18-20] from the point of view of complex analysis.

### 5.1 Potentially-solenoidal vector field

Let us consider the function

$$u(x, y) = \phi = \mathrm{angle}(x, y) = \arg(z),$$
$$z = x + iy = |z| e^{i\phi}, \quad (1)$$

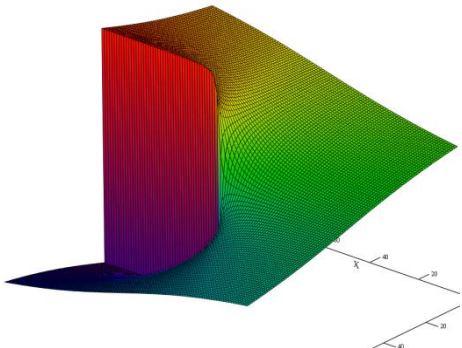

Fig. 7 Function $u(x, y)$ surface

the graph of which is shown in Fig.7. Function $u(x, y)$ is not continuous at $x \in [0, +\infty)$. Let us calculate the gradient of function $u(x, y)$ using the polar (cylindrical) coordinate system, then we obtain

$$\nabla u(x, y) = \vec{e}_\rho \frac{\partial \phi}{\partial \rho} + \vec{e}_\phi \frac{1}{\rho}\frac{\partial \phi}{\partial \phi} = \frac{\vec{e}_\phi}{\rho}. \quad (2)$$

On the one hand the vector field (2) is a potential, however, the rotor (curl) of such a field is not equal to zero. Indeed, let us calculate the circulation of the field (2) on a circle $C_R$ of radius $R$ with the center in the origin, and we obtain

$$\oint_{C_R} (\nabla u, d\vec{\rho}) = \int_0^{2\pi} \frac{1}{R} R d\phi = 2\pi, \qquad (3)$$

were $d\vec{\rho} = \vec{e}_\phi R d\phi$. Difference of the potential field rotor (2) from zero is connected with non-smoothness of function $u(x,y)$ (1). Function (1) is not unique and when making full turns around the origin, there is a gap in values $\phi = 0$ and $\phi = 2\pi$. At the origin function $u(x,y)$ is not defined. If we perform a double bypass, the difference between the initial and final value will be $4\pi$. At $k$-times bypass it will be $2\pi k$, respectively. If function $u(x,y)$ was smooth, then integral (3) would be equal to zero.

From a physical point of view the equality of integral (3) to zero has the interpretation of work performed by the force $\nabla u$. For example, for a potential force field with sufficient smoothness, as it is known, work equals zero. However, for non-smooth potentials (1) «work» in a closed circle is different from zero. Analogous to a non-zero «work» (3) from a physical point of view is the circulation of the magnetic field around a conductor with a current in a certain circuit $C$ containing a current-carrying conductor. In this case, the circulation will be equal to current:

$$\text{rot}\, \vec{H} = \vec{J},$$
$$\oint_C (\vec{H}, d\vec{\rho}) = I. \qquad (4)$$

Note that for the magnetic field, the potential may be a function of (1), which will correspond to a unit current perpendicular to the axis XOY

$$\vec{H} = \frac{I}{2\pi} \frac{1}{\rho} \vec{e}_\phi. \qquad (5)$$

From the point of view of the complex analysis the function $\vec{H}$ has a pole of first order at the origin, therefore, it is not analytic inside of closed contour $C$, and the contour integral is known to be expressed through the residue of function (5).

### 5.2 Motion along the closed orbit

Let us calculate the integral of momentum $\langle \vec{p} \rangle$ along closed path $C$, considering the decomposition (3), (4) §1:

$$\oint_C (\langle \vec{p} \rangle, d\vec{r}) = m \oint_C (\langle \vec{v} \rangle, d\vec{r}) = -\alpha m \oint_C (\nabla \Phi, d\vec{r}) + m\gamma \oint_C (\vec{A}, d\vec{r}) =$$
$$= -\alpha m \Phi \Big|_1^1 + m\gamma \int_\Sigma (\text{rot}\, \vec{A}, d\vec{\sigma}), \qquad (6)$$

where $\Sigma$ – the surface stretched on the contour $C$. Let us present scalar potential $i\Phi$ in the form of function $\operatorname{Ln}\left[\dfrac{\Psi}{\bar{\Psi}}\right]$ (9) §1 which is multivalent (see Fig.2), therefore

$$-\alpha m\Phi\Big|_1^1 = \dfrac{\hbar}{2m}m\Phi\Big|_1^1 = \dfrac{\hbar}{2}\Phi\Big|_1^1 = \dfrac{\hbar}{2}(2\varphi+2\pi k)\Big|_1^1 = \hbar\varphi\Big|_1^1 = \hbar\varphi\Big|_0^{2\pi k} = \hbar 2\pi k = hk,$$

where $h \stackrel{\text{det}}{=} 2\pi\hbar$, $k \in \mathbb{Z}$. As a result, expression (6) takes the form

$$\oint_C (\langle\vec{p}\rangle, d\vec{r}) = hk - e\int_\Sigma (\vec{B}, d\vec{\sigma}). \tag{7}$$

Let us consider the integral of the potential component of the momentum. Using the Legendre transformation, we obtain:

$$\oint_C (\langle\vec{p}_p\rangle, d\vec{r}) = \int_0^T (\langle\vec{v}\rangle, \langle\vec{p}_p\rangle) dt = \int_0^T \tilde{L}(\vec{r}, \langle\vec{v}\rangle, \langle\vec{v}_s\rangle, t) dt + \int_0^T \tilde{H}(\vec{r}, \langle\vec{p}_p\rangle, \langle\vec{p}_s\rangle t), \tag{8}$$

where $d\vec{r} = \langle\vec{v}\rangle dt$ on the trajectory; $T$ – time (period) of passing of contour $C$. From a mathematical point of view, the integral of $\tilde{L}$ is action $\tilde{S}$, however, in (8) the trajectory is closed. Replacing the integral of $\tilde{L}$ by action $\tilde{S}$, and considering ratio (17) §1, we obtain

$$\oint_C (\langle\vec{p}_p\rangle, d\vec{r}) = \tilde{S} + \int_0^T \tilde{H}(\vec{r}, \langle\vec{p}_p\rangle, \langle\vec{p}_s\rangle, t) dt = \hbar(\varphi_T - \varphi_0) + H_0 T = \\ = \hbar 2\pi k + H_0 T = hk + H_0 T, \tag{9}$$

where $H_0$ – the average value $\tilde{H}(\vec{r}, \langle\vec{p}_p\rangle, \langle\vec{p}_s\rangle, t)$ on interval $(0,T)$.

Comparing expressions (7) and (9) in the case of the irrotational velocity field ($\vec{A} = \vec{0}$) we can see their difference in value $\int_0^T \tilde{H}(\vec{r}, \langle\vec{p}_p\rangle, \langle\vec{p}_s\rangle, t) dt$, although, from general considerations, (7) and (9) shall give the same result. This contradiction appears because of the assumptions made in the derivation of expression (7). In (7) the following statement was supposed as completed one:

$$\int_1^2 (\nabla\Phi, d\vec{r}) = \Phi\Big|_1^2, \tag{10}$$

which in the general case for function $\Phi$ as for the function of two variables $\vec{r}$ and $t$, which are changed simultaneously with the movement along the contour ($\vec{r} = \vec{r}(t)$) is incorrect. Statement (10) is valid, if $\Phi$ is a function only of variable $\vec{r}$, i.e. $\Phi = \Phi(\vec{r})$ or if $\vec{r}$ and $t$ are independent variables. In this case

$$d\Phi(\vec{r}) = (\nabla\Phi, d\vec{r}),$$

and expression (10) is valid. If $\Phi = \Phi(\vec{r}, t)$ and $t$ are dependent and $(\vec{r} = \vec{r}(t))$ then the differential has a view

$$d\Phi(\vec{r}(t), t) = \frac{\partial \Phi}{\partial t} dt + (\nabla \Phi, d\vec{r}) = -\frac{2}{\hbar} \tilde{H} dt + (\nabla \Phi, d\vec{r}),$$

$$(\nabla \Phi, d\vec{r}) = d\Phi(\vec{r}(t), t) + \frac{2}{\hbar} \tilde{H} dt,$$

which gives expression (9).

The obtained relations (7) and (9) are similar to the well-known of Bohr-Sommerfeld quantization rule:

$$\oint p \, dx = hk, \qquad \text{or} \qquad \oint p \, dx = hk + \frac{h}{2}, \qquad \text{or} \qquad \oint p_i \, dq_i = hk_i + \gamma_i. \qquad (11)$$

where the integral is taken over the period of change of the generalized coordinate $q_i$, $\gamma_i$ is the number of order of unity, depending on the nature of the boundary conditions for the given degrees of freedom.

Integrals such as (7), (9), (11) represent the volume of phase space corresponding to trajectories of particles, and indicate that this quantity is quantized, that is, there is a minimum step (Planck constant) of such change.

## §6 Examples of quantum systems

Let us consider two examples illustrating the results obtained above. In both examples the vortex field of the flow of probabilities of (2) §5 was used, and a function of the probability density is stationary $f = f(r)$.

### 6.1 Motion in a central field

Let us consider an extended version of function (1) §5

$$\varphi = c_1 u + c_2 = c_1 \phi + c_2, \qquad (1)$$

where $c_1, c_2$ are the values that do not depend on coordinates $x, y, z$; $\phi$ − azimuth angle. For cylindrical coordinate systems we will use the notations $(\rho, \phi, z)$ and for spherical − $(r, \theta, \phi)$. Functions $\varphi$ and $u$ (1) are partial solutions of nonlinear equations of a divergent type that is included in the magnetostatic problems [9,10]

$$\text{div}\left[\mu(|\nabla u|) \nabla u\right] = 0, \qquad (2)$$

where $\mu$ is the magnetic permeability. Function (1) is a solution to (2) with any dependence of $\mu(H)$ on the field $\vec{H} = -\nabla u$ (this can be verified by a direct substitution of (1) into (2)). In papers [9,10] the construction of solutions of the type (1), having an unlimitedly growing $|\nabla u|$ in the corner domain was considered. The degree of growth $|\nabla u|$ was $\sim \rho^{\lambda-1}$, where $\lambda < 1$.

Continuity equation (1) §1 in the stationary case $f(\vec{r},t) = f(\vec{r})$ goes into (2), and function $\mu$ corresponds to function $f(\vec{r}) = \Psi\bar{\Psi}$, and vector field $\nabla u$ corresponds to the field of the flow probabilities velocity $\langle \vec{v} \rangle$. Let us select the following values as $c_1$ and $c_2$

$$c_1 = k \in \mathbb{Z}, \; c_2 = -\frac{E}{\hbar}t = -\beta E t, \qquad (3)$$

then in accordance with (3) §1 (1), (3) we obtain

$$\langle \vec{p} \rangle = m \langle \vec{v} \rangle = -\alpha m \nabla_r \Phi = -2m\alpha \nabla_r \varphi = \hbar \nabla_r \varphi = \hbar \nabla_r \varphi = -i\hat{p}\varphi = \frac{\hbar k}{\rho} \vec{e}_\phi, \qquad (4)$$

$$\nabla_r \tilde{S} = \nabla_r \hbar \varphi = \hbar \nabla_r \varphi = -i\hat{p}\varphi = \langle \vec{p}_p \rangle = \langle \vec{p} \rangle,$$

where $\hat{p} = i\hbar \nabla_r$, $\rho = r \sin \theta$. From (4) it follows that the expression for the angular momentum $\vec{M} = [\vec{\rho}, \langle \vec{p} \rangle]$ takes the form

$$\vec{M} = [\vec{\rho}, \langle \vec{p} \rangle] = \left[ \rho \vec{e}_r, \frac{\hbar k}{\rho} \vec{e}_\phi \right] = \hbar k \vec{e}_z. \qquad (5)$$

From (5) we can see that the module of the angular momentum $M = \hbar k, \; k \in \mathbb{Z}$, and its direction is parallel to axis OZ. Thus, the angular momentum $\vec{M}$ is perpendicular to the plane of rotation, and its module according to (5) is quantized. The same ratio for the angular momentum exists in quantum mechanics, for the projection of angular momentum on the axis $OZ$ with magnetic field line $L_z = \hbar k$, where $k$ – the magnetic quantum number. From the point of view of the complex analysis, number $k$ defines the valent number of the Riemann surface.

As was shown in §1, $\varphi$ is the phase of the wave function, the velocity potential of the flow of probability and action. Velocity $\langle \vec{v} \rangle$ in this case is not explicitly divided into potential and vortex component according to the Helmholtz theorem, as the potential (1) is not a smooth function, hence, the theorem cannot be applied. In representation (4) the nucleus of an atom from a mathematical point of view is in a special point – in the pole of the velocity first order. Integrals (7), (9) §5 (4) will take the form

$$\oint_C (\langle \vec{p} \rangle, d\vec{\rho}) = hk. \qquad (6)$$

Expression (6) is also consistent with the Bohr-Sommerfeld quantization rule (11) §5. The orbit will correspond to different Riemann surface Fig.2.

As considering condition (4) according to [9-10] equation (2) will be fulfilled for any function of probability density $f = f(r)$, as an example, we are going to consider $f(r) = |\Psi|^2 = Cr^\upsilon e^{-\kappa r}$, where constant $C$ is determined from the condition of normalization

$$C \int_0^{2\pi} d\phi \int_0^\pi \sin\theta d\theta \int_0^{+\infty} r^{\upsilon+2} e^{-\kappa r} dr = \frac{4\pi C}{\kappa^{\upsilon+3}} \int_0^{+\infty} t^{\upsilon+2} e^{-t} dt = \frac{4\pi C}{\kappa^{\upsilon+3}} \Gamma(\upsilon+3) = 1,$$

$$C = \frac{\kappa^{\upsilon+3}}{4\pi \Gamma(\upsilon+3)}, \tag{7}$$

where $\Gamma$ is the gamma function. Thus, due to (1), (3), (7) the wave function is

$$\Psi(r,t) = \frac{\kappa^{\upsilon+3}}{4\pi \Gamma(\upsilon+3)} r^{\upsilon} e^{-\kappa r} e^{i\left(k\phi - \frac{E}{\hbar}t\right)} = \Psi_0(\vec{r}) e^{-i\frac{E}{\hbar}t}, \tag{8}$$

where $\Psi(\vec{r},0) = \Psi_0(\vec{r}) \in \mathbb{C}$. Potential energy $U$ (10A) §1 takes the form

$$U(\vec{r}) = -\frac{1}{\beta}\left\{\frac{\partial \varphi}{\partial t} + \alpha\left(\frac{\Delta|\Psi|}{|\Psi|} - |\nabla\varphi|^2\right)\right\} = E + \frac{\hbar^2}{2m}\left(\frac{\Delta|\Psi|}{|\Psi|} - |\nabla\varphi|^2\right) =$$

$$= E + \frac{\hbar^2}{2m}\left(\kappa^2 + \frac{\upsilon(\upsilon+1)}{r^2} - \frac{2\kappa(\upsilon+1)}{r} - \frac{k^2}{r^2 \sin^2\theta}\right),$$

$$U(\vec{r}) = E_0 + \frac{a(\theta)}{r^2} - \frac{b}{r}, \tag{9}$$

where

$$E_0 = E + \frac{\hbar^2 \kappa^2}{2m}, \quad a(\theta) = \frac{\hbar^2}{2m}\left[\upsilon(\upsilon+1) - \frac{k^2}{\sin^2\theta}\right], \quad b = \frac{\hbar^2 \kappa(\upsilon+1)}{m}.$$

In the particular case when $E = -\frac{\hbar^2 \kappa^2}{2m}$ potential (9) takes the form [2]:

$$U(\vec{r}) = \frac{a}{r^2} - \frac{b}{r}, \tag{10}$$

where $a,b$ are positive constants. Substituting the wave function (8) and potential (9) into the Schrödinger equation (6) §1, we obtain the right identity

$$i\hbar \frac{\partial \Psi}{\partial t} = -\frac{\hbar^2}{2m}\Delta\Psi + U(\vec{r})\Psi, \tag{11}$$

indeed

$$i\hbar \frac{\partial \Psi}{\partial t} = i\hbar C r^{\upsilon} e^{-\kappa r} e^{i\varphi} i \frac{\partial \varphi}{\partial t} = -\hbar C r^{\upsilon} e^{-\kappa r} e^{i\varphi}\left(-\frac{E}{\hbar}\right) = E\Psi$$

$$-\frac{\hbar^2}{2m}\Delta\Psi = -C\frac{\hbar^2}{2m}\left(\frac{e^{i\varphi}}{r^2}\frac{\partial}{\partial r} r^2 \frac{\partial}{\partial r}\left(r^{\upsilon} e^{-\kappa r}\right) + \frac{r^{\upsilon} e^{-\kappa r}}{r^2 \sin^2\theta}\frac{\partial^2 e^{i\varphi}}{\partial \phi^2}\right) =$$

$$= -\frac{\hbar^2}{2m}\left(\kappa^2 - \frac{2\kappa(\upsilon+1)}{r} + \frac{\upsilon(\upsilon+1)}{r^2} - \frac{k^2}{r^2 \sin^2\theta}\right)\Psi,$$

$$U(\vec{r})\Psi = E\Psi + \frac{\hbar^2}{2m}\left(\kappa^2 + \frac{\upsilon(\upsilon+1)}{r^2} - \frac{2\kappa(\upsilon+1)}{r} - \frac{k^2}{r^2 \sin^2\theta}\right)\Psi.$$

Classical potential $e\chi$ as in (10A) §1 has the view

$$e\chi = U - \frac{\hbar^2}{2m}\frac{\Delta|\Psi|}{|\Psi|} = E - \frac{\hbar^2 k^2}{2mr^2 \sin^2\theta}. \qquad (12)$$

The total energy $W$ (10) §1, taking (4) into account, (12) is a constant value:

$$-\frac{\hbar}{2}\frac{\partial \Phi}{\partial t} = E = W(\vec{r}) = W(\rho) = \frac{\hbar^2 k^2}{2mr^2 \sin^2\theta} + e\chi = E = const. \qquad (13)$$

The solution of equation (1), (2) §1 with a velocity of (4) can be found in another way. As the velocity is representable in the form of (4), $Q = \text{div}_r \langle \vec{v} \rangle = 0$ then equation (1), (2)§1 will take the form of transport equation (2A) §1. The solution of equation (2A) §1 can be found by the method of characteristics. Along the characteristics defined by the equation

$$\frac{d\vec{\rho}}{dt} = -\langle \vec{v} \rangle = -\frac{\hbar}{m}\frac{k}{\rho}\vec{e}_\phi \qquad (14)$$

Function $f(\vec{\rho}(t),t) = const$. Making substitutions

$$\vec{e}_\phi = -\vec{e}_x \sin\phi + \vec{e}_y \cos\phi, \ \vec{\rho} = \vec{e}_x x + \vec{e}_y y, \ x = \rho\cos\phi, \ y = \rho\sin\phi,$$

Equation (14) takes the form

$$\begin{cases} \dfrac{dx}{dt} = \dfrac{\hbar}{m}\dfrac{k}{\rho}\sin\phi = \dfrac{\hbar k}{m}\dfrac{y}{\rho^2}, \\ \dfrac{dy}{dt} = -\dfrac{\hbar}{m}\dfrac{k}{\rho}\cos\phi = -\dfrac{\hbar k}{m}\dfrac{x}{\rho^2}. \end{cases} \qquad (15)$$

Dividing equation (15), we obtain

$$\frac{dx}{dy} = -\frac{y}{x},$$

$$xdx = -ydy,$$

$$\left.\frac{x^2}{2}\right|_{x_0}^{x} = -\left.\frac{y^2}{2}\right|_{y_0}^{y},$$

$$x^2 + y^2 = x_0^2 + y_0^2 \overset{det}{=} \rho_0^2 = const. \qquad (16)$$

Equation (16) sets the equation of the circle with the radius of $\rho_0$, therefore, the characteristics of (14) have the form of concentric circles along which $f(\vec{\rho}(t),t) = const$. It turns out that equations (1), (2) §1 have a stationary solution and are stationary, i.e. have the form

$$(\langle \vec{v} \rangle, \nabla_\rho f) = 0. \qquad (17)$$

From equation (17) it also follows that $\nabla_\rho f \perp \langle \vec{v} \rangle$, $\nabla_r f$ not a zero vector in the general case, and vector $\langle \vec{v} \rangle$ according to (18) is also non-zero. As vector $\langle \vec{v} \rangle$ has only $\vec{e}_\phi$ component, then vector $\nabla_\rho f$ due to (17) has only radial component $\vec{e}_\rho$, that is, $f$ depends only on the radius $f = f(\rho)$ that was required to show.

As the solution of the stationary equation of continuity (2) under conditions (4) admits solutions in the form of characteristics that are circles along which the probability density $f = const$, evolution operator (5) §4 becomes operator (34) §3, which produces a rotation in the complex plane domain $D_{\Psi_1}$ to angle $\Delta\varphi = \varphi_2 - \varphi_1$.

### 6.2 The Dirac string model

In the previous example, velocity $\langle \vec{v} \rangle$ according to (3) §1 was presented in the form of $-\alpha \nabla \Phi$. Thus the obtained field $\langle \vec{v} \rangle$ (4) is vortex. On the other hand according to (3) §1 the vortex component of the field $\langle \vec{v} \rangle$ is $\gamma \vec{A}$, therefore, according to (3),(4), we can assume that

$$\langle \vec{v}_p \rangle = -\alpha \nabla \Phi = 0, \ \nabla_r \tilde{S} = \langle \vec{p}_p \rangle = 0,$$
$$\langle \vec{v}_s \rangle = \gamma \vec{A} = \frac{\hbar k}{mr \sin \theta} \vec{e}_\phi, \quad (18)$$
$$\langle \vec{v} \rangle = \langle \vec{v}_s \rangle.$$

From (18) it follows that the vector potential has the form:

$$\vec{A} = -\frac{\hbar k}{q_e r \sin \theta} \vec{e}_\phi, \quad (19)$$

where the electrical charge is indicated as $q_e$. As the scalar velocity potential according to (1), (3) at $c_1 = 0$, we take

$$\Phi = 2c_2 = -\frac{2}{\hbar} E_k t. \quad (20)$$

*Comment*

Here, as in the previous example, the Bohr-Sommerfeld quantization rule takes place (6). In the particular case of the Coulomb potential we can get energy and the radii of the corresponding Bohr model of the atom. Indeed the total energy $W$ (10) §1 has the view

$$W(\vec{r}) = \frac{\hbar^2 k^2}{2mr^2 \sin^2 \theta} - \frac{Ze^2}{4\pi\varepsilon_0 r}. \quad (21)$$

Energy (21) will depend only on the radius $r$, if the plane in which the trajectory lies, passes through the origin. Therefore, without limiting generality, let us consider $\theta = \frac{\pi}{2}$, then $r = \rho$

$$W(\rho) = W(r) = \frac{\hbar^2 k^2}{2mr^2} - \frac{Ze^2}{4\pi\varepsilon_0 r}. \tag{22}$$

The energy (22) remains constant $W(\vec{r}) = E = const$ if the radius is fixed. Since energy (22) is constant along a circular path $\vec{r} = \vec{r}(t)$ then $\frac{d}{dt}W(\vec{r}(t)) = 0$:

$$0 = \frac{\partial W}{\partial t} + (\langle\vec{v}\rangle, \nabla W) = \frac{\hbar k}{mr}\frac{\partial W}{\partial r}(\vec{e}_\phi, \vec{e}_r). \tag{23}$$

Equation (23) is presented, as $(\vec{e}_\phi, \vec{e}_r) = 0$, however, according to the equation of Hamilton-Jacobi (10) §1, along the trajectory $\nabla W = \vec{\theta}$, we obtain

$$-\nabla\frac{\partial \Phi}{\partial t} = \nabla E = \vec{\theta} = \frac{2}{\hbar}\nabla W. \tag{24}$$

From condition (24) the expressions for the radii of orbits in Bohr model of the atom are presented

$$\frac{\partial W}{\partial r} = -\frac{\hbar^2 k^2}{mr^3} + \frac{Ze^2}{4\pi\varepsilon_0 r^2} = 0,$$

$$r_k = \frac{4\pi\varepsilon_0 \hbar^2 k^2}{Ze^2 m}. \tag{25}$$

Substituting (25) into (22) we obtain the known energy levels, corresponding to the radii $r_k$:

$$E_k = -\frac{Z^2 e^4 m}{32\pi^2 \varepsilon_0^2 \hbar^2 k^2}.$$

Thus, let us consider the stationary continuity equation (2) with flow probabilities (18). The function of density of probability will remain unchanged $f(r)$, i.e. $|\Psi| = |\tilde{\Psi}|$. Consequently, a new wave function $|\Psi| = |\tilde{\Psi}|$ according to (18), (20) takes the form

$$\tilde{\Psi} = Cr^\upsilon e^{-\kappa r} e^{-i\frac{E}{\hbar}t}, \tag{26}$$

where $C$ has the form (7). Substituting (19) and (20) into (7), (10A) §1 we obtain the expression for potentials $\tilde{U}$ and $q_e\tilde{\chi}$, respectively

$$\tilde{U}(\vec{r}) = E + \frac{\hbar^2}{2m}\frac{\Delta|\tilde{\Psi}|}{|\tilde{\Psi}|} = E + \frac{\hbar^2 \kappa^2}{2m} + \frac{\hbar^2}{2m}\frac{\upsilon(\upsilon+1)}{r^2} - \frac{\hbar^2}{m}\frac{\kappa(\upsilon+1)}{r},$$

$$\tilde{U}(\vec{r}) = E_0 + \frac{\tilde{a}}{r^2} - \frac{b}{r}, \qquad (27)$$

where $\tilde{a} = \dfrac{\hbar^2 \upsilon(\upsilon+1)}{2m}$, and $E_0$, $b$ are defined in (9).

$$q_e \tilde{\chi} = -\frac{\hbar^2 k^2}{2mr^2 \sin^2\theta} - \frac{\hbar^2}{2m} \frac{\Delta|\tilde{\Psi}|}{|\tilde{\Psi}|} + \tilde{U} = E - \frac{\hbar^2 k^2}{2mr^2 \sin^2\theta}, \qquad (28)$$

The potential (27) is different from that of the potential (9), considered in the previous example. The Schrödinger equation (6) §1 takes the form

$$i\hbar \frac{\partial \tilde{\Psi}}{\partial t} = \frac{\hat{p}^2}{2m} \tilde{\Psi} - \frac{q_e}{m}(\vec{A},\hat{p})\tilde{\Psi} + \tilde{U}\tilde{\Psi}. \qquad (29)$$

Substituting wave function (26) into equation (29), we obtain the right identity:

$$i\hbar \frac{\partial \tilde{\Psi}}{\partial t} = -i\frac{E}{\hbar} i\hbar C r^\upsilon e^{-\kappa r} e^{-i\frac{E}{\hbar}t} = E\tilde{\Psi},$$

$$\frac{\hat{p}^2}{2m}\tilde{\Psi} = -\frac{\hbar^2}{2m} C e^{-i\frac{E}{\hbar}t} \Delta(r^\upsilon e^{-\kappa r}) = -\frac{\hbar^2}{2m}\tilde{\Psi}\left(\kappa^2 - \frac{2\kappa(\upsilon+1)}{r} + \frac{\upsilon(\upsilon+1)}{r^2}\right)$$

$$\frac{q_e}{m}(\vec{A},\hat{p})\tilde{\Psi} = \frac{i\hbar^2 k}{q_e r \sin\theta} \frac{q_e}{m} C e^{-i\frac{E}{\hbar}t} \frac{\partial}{\partial r}(r^\upsilon e^{-\kappa r})(\vec{e}_\phi,\vec{e}_r) = 0$$

$$\tilde{U}\tilde{\Psi} = \left[E + \frac{\hbar^2 \kappa^2}{2m} + \frac{\hbar^2}{2m}\frac{\upsilon(\upsilon+1)}{r^2} - \frac{\hbar^2}{m}\frac{\kappa(\upsilon+1)}{r}\right]\tilde{\Psi}$$

Let us find a vector field of magnetic induction $\vec{B}$ as the rotor of the vector potential $\vec{A}$ (19). Vector field $\vec{A}$ has a pole of first order on axis OZ. Finding $\operatorname{rot}\vec{A}$ for $\rho \neq 0$, we obtain

$$\vec{B} = \operatorname{rot}\vec{A} = -\frac{\hbar k}{q_e}\operatorname{rot}\frac{\vec{e}_\phi}{\rho} = -\frac{\hbar k}{q_e}\frac{\vec{e}_z}{\rho}\frac{\partial}{\partial \rho}\left(\rho\frac{1}{\rho}\right) = \vec{0}. \qquad (30)$$

To calculate $\vec{B}$ on axis OZ we use the Stokes formula. Let $C_R$ be a circle of radius $R$ with the center at axis OZ ($\rho = 0$), and $\Sigma$ a flat surface stretched on contour $C_R$. Let us calculate the flux of vector $\vec{B}$ through the surface $\Sigma$, then we obtain

$$\int_\Sigma \vec{B}d\vec{\sigma} = \int_\Sigma \operatorname{rot}\vec{A}d\vec{\sigma} = \oint_{C_R} \vec{A}d\vec{l} = -\frac{\hbar k}{q_e}\int_0^{2\pi}\frac{1}{R}Rd\phi = -\frac{2\pi\hbar k}{q_e}. \qquad (31)$$

Let us consider the Dirac delta function $\delta^2(\rho,\phi) = \dfrac{\delta(\rho)}{2\pi\rho}$, having the property

$$\int_\Sigma \delta^2(\rho,\phi)d\sigma = \int_0^{2\pi} d\phi \int_0^R \frac{\delta(\rho)}{2\pi\rho}\rho d\rho = 1. \qquad (32)$$

Comparing (30)-(32) for $\vec{B}$ we obtain the expression

$$\vec{B} = -\frac{2\pi\hbar k}{q_e}\frac{\delta(\rho)}{2\pi\rho}\vec{e}_z = -\frac{q_m^{(Wb)}\delta(\rho)}{2\pi\rho}\vec{e}_z, \qquad (33)$$

$$q_m^{(Wb)} \stackrel{\text{det}}{=} \frac{2\pi\hbar k}{q_e}. \qquad (34)$$

or

$$\vec{H} = -\frac{2\pi\hbar k}{q_e\mu_0}\frac{\delta(\rho)}{2\pi\rho}\vec{e}_z = -\frac{q_m^{(A\cdot m)}\delta(\rho)}{2\pi\rho}\vec{e}_z, \qquad (33A)$$

$$q_m^{(A\cdot m)} \stackrel{\text{det}}{=} \frac{2\pi\hbar k}{q_e\mu_0}, \qquad (34A)$$

Value $q_m$ is known as a magnetic charge [30-32]. In the system SI the measurement unit of magnetic charge $q_m^{(Wb)}$ is «Wb», and of $q_m^{(A\cdot m)}$ «A·m». The relation $q_m^{(Wb)} = \mu_0 q_m^{(A\cdot m)}$ is valid. From (34), (34A) there is the condition of the Dirac quantization of the magnetic charge $q_m$ [33]

$$\frac{q_e q_m^{(Wb)}}{2\pi\hbar} = k \in \mathbb{Z}, \quad \frac{q_e q_m^{(A\cdot m)}}{2\pi\varepsilon_0 \hbar c^2} = k \in \mathbb{Z}. \qquad (35)$$

Magnetic field (33), (33A) corresponding to the magnetic charge $q_m$ has a single field line coinciding with axis OZ. Note that this system is a dipole that is not a Dirac monopole.